# Direct evidence for electron hole depletion in $La_{0.5}Sr_{0.5}FeO_{3-\delta}$ upon B-site cation doping

*Artur Braun*[1], *Defne Bayraktar* [1], *Ashley S. Harvey*[2], *Daniel Beckel*[2], *John A. Purton*[3]

*Peter Holtappels*[1], *Ludwig J. Gauckler*[2], *Thomas Graule*[1]

[1]Laboratory for High Performance Ceramics

EMPA – Swiss Federal Laboratories for Materials Testing & Research

CH – 8600 Dübendorf, Switzerland

[2]Department of Nonmetallic Materials

Swiss Federal Institute of Technology

CH – 8053 Zürich, Switzerland

[3]Synchrotron Radiation Source

Daresbury Laboratory, WA4 4AD Warrington, UK

Abstract

Partial substitution of Fe by Ta in the hole-doped $La_{0.5}Sr_{0.5}FeO_3$ decreases the electric conductivity by up to three orders of magnitude. This decrease is in immediate correlation with a decrease of the electron hole concentration and a shift of the spectral weight within the O(1s)-Fe(3d) mixed states from $e_g$ bands near the Fermi energy to $t_{2g}$ bands. Corresponding differences in the Fe(2p) and O(1s) X-ray absorption spectra reveals formation of states with increased covalency in the initial state and states which contain O(2p) character, and that hole states are responsible for transport changes in the material. The intensity ratio of $e_g$ and $t_{2g}$ bands appears to be a spectral indicator for hole formation.

The rare earth transition metal oxides display a manifold of technologically relevant functionalities, including such spectacular phenomena such as high temperature superconductivity, colossal magnetoresistivity and metal-insulator transitions. Electron hole doping aims at improving the high temperature conductivity of such oxides for applications in solid oxide fuel cells or oxygen membranes in catalysis. LaSrFe-oxide is such a system that is increasingly under scrutiny, and adequate theoretical description of its behaviour is still lacking.
The heterovalent substitution of $La^{3+}$ in the charge transfer insulator $LaFeO_3$ with $Sr^{2+}$ creates electron hole states with substantial O(2p) character near the Fermi level [1-4]. In addition to the hole doping effect, the increasing radius of the A-site ion from La to Sr decreases the typical rhombohedral lattice distortion, which in turn increases the hybridization of the O(2p)-Fe(3d) states, and thus the one-electron band width W [5], which is a function of the super-exchange angle θ, formed by the Fe-O-Fe configuration: $W \sim \cos^2\theta$ [6].

The electric conductivity of $La_{1-x}Sr_xFeO_3$ (LSF) increases with increasing x, with $SrFeO_3$ displaying typical metal conductivity [3,5]. During hole doping, La and Sr are considered as retaining their formal valence of $La^{3+}$ and $Sr^{2+}$. Charge balance is then maintained by adjustment of the oxidation state of Fe and by onset of an oxygen deficiency "δ", which is also increasing with x. While the formal valence is $Fe^{3+}$ in $LaFeO_3$ and $Fe^{4+}$ in $SrFeO_3$, the actual relative $Fe^{4+}$ concentration in $SrFeO_3$ is only about 40% or even less, due to a substantial value of "δ", particularly in $SrFeO_{3-\delta}$. With every $Fe^{4+}$ replacing an $Fe^{3+}$ in LSF, one electron hole is created which can serve as charge carrier for electronic transport [7]. Our starting reference material is LSF with x = 5 (LSF50), which has the maximum oxygen, electron hole and electron conductivity [4,8].

Phase pure LSF and substituted LSF samples with R-3c symmetry as listed in the Table were synthesized by solid state reaction and their conductivity was measured with the 4-point technique on sintered bars [9]. More details on LSF, including crystal structure and thermal expansion, thermochemical behaviour, and electric and magnetic properties are available in [4,9-12]. The stoichiometry of LSF50 imposes a formal valency of 3.5 for iron, with $Fe^{3+}$ and $Fe^{4+}$ being present to equal amounts in the unit cell. The formal hole concentration $\tau = Fe^{4+}/ Fe^{3+}$ is therefore $\tau=1$. The gravimetrically obtained oxygen deficiency served to calculate the actual hole concentration $\tau$ (Table).

The conductivity of the four samples in air as a function of temperature is shown in Figure 1. Just by substituting the B-site cation from Fe to Ta by 20%, the conductivity varies by three orders of magnitude, so that it is necessary to plot the conductivity on a logarithmic scale. LSF50 has the overall highest conductivity at 300 K (50 S/cm) and also at higher temperatures (380 S/cm), systematically followed by LSFTi20 (1.5 S/cm, 100 S/cm), LSFTa10 (0.7 S/cm, 140 S/cm), and LSFTa20, which has the lowest conductivity (0.006 S/cm, 20 S/cm). Apart from the absolute conductivity value, the variation profile vs. temperature is similar for all samples. The initial convex increase is due to polaron activation and is followed at around 500 K - 800 K by a conductivity maximum or a plateau. The conductivities of LSFTi and LSFTa10 show close similarity, but LSFTa10 has a smaller low temperature conductivity. Comparison of $\sigma$ and $\tau$ in the table suggests a correlation between conductivity and substitution level.

Soft x-ray absorption spectra were recorded at 300 K at beamline 1.1. at Daresbury Laboratory Synchrotron Radiation Source, United Kingdom. The vacuum recipient pressure was 5·E-10 torr. The four panels in Figure 2 display calibrated and normalized Fe (2p) absorption spectra of our samples. The multiplet structure of 3d metals such as iron and its x-ray spectroscopic assessment are generally well understood [2,13,14] and provide a unique probe for the local spin moment of magnetic atoms [15]. The degeneracy of the Fe (2p) core hole level is lifted by the spin-orbit coupling, resulting in the $2p_{3/2}$ and $2p_{1/2}$ multiplets ($L_3$ and $L_2$ absorption edges, respectively) at around 708 and 723 eV. The centroids of L3 and L2 are separated by an exchange energy of $\Delta \approx 13$ eV (Table). The octahedral crystal field lifts the degeneracy of the $2p_{3/2}$ and $2p_{1/2}$ levels so that two levels with $t_{2g}$ and $e_g$ symmetry are created, as indicated by the two structures at 707 eV and 708.5 eV, and at 720.5 eV and 723 eV, respectively.
The chemical shift of the white lines of the spectra reveals that the average oxidation state of the Fe differs from sample to sample. The white line with the lowest energy is that of LSFTa20, 708.8 eV, followed by LSFTi20 (709 eV), LSFTa10 (710 eV), and LSF50 (710 eV). This sequence is reasonable, because LSFTa20 contains only $Fe^{3+}$, thus having the lowest Fe oxidation state and no holes. The relative hole concentration is $\tau$= 0, 0.25, 0.28, and 0.52 in the aforementioned series. LSFTa20 with only high spin $Fe^{3+}$ in $3d^5$ configuration with parallel spins has the largest multiplett branching ratio $Q = L_3/( L_2+L_3) = 0.70$ (Table 1) [16]. Due to the larger cation size for $Ti^{4+}$ and $Ta^{5+}$ compared to $Fe^{4+}$, the unit cell of LSF is expanding upon substitution with Ti and Ta [9], possibly implying that the remaining Fe ions have more space available allowing for parallel spins in the Fe.
In line with previous conclusions on LSF [2], we find that the Fe ions in our samples remain essentially in the ferromagnetic high spin $3d^5$ configuration ($t_{2g}^3$ $e_g^2$), and identify the spectrum of LSF50 in Figure 1 in reference [2] as the one that resembles most similarities with our spectra.

To better understand the transport mechanism of the B-site doped LSF, we recorded O(1s) XAS spectra, which represent to a good approximation the unoccupied oxygen p projected density of states (DOS) [17]. Since the other orbitals are strongly hybridized with the O(2p) orbitals, O(1s) XAS reflects also the empty Fe(3d) and La(5d) bands. In the O(1s) spectra in Figure 3 normalized to unity at 560 eV, we study the pre-edge features between 526 and 530 eV. They show systematic differences depending on the doping situation. The pre-peaks at 527.7 eV and 529.3 eV represent mixed O(2p)-Fe(3d) states subject to the ligand field splitting of $\Delta \approx 1.3$ eV in the octahedral $FeO_6$ environment, and hence form $e_g$ and $t_{2g}$ states in the band gap [2,3,18]. Note that we do not find a peak shift in our spectra. Instead, we notice a variation of the peak intensities. The structures at 534.4 and 536 eV are due to O (2p) states covalently mixed with La (5d) and Sr (4d) bands. The structures at 540.0 eV and 541.8 eV are due to interactions of the oxygen with the Fe (4sp) and the La (6sp) bands [19,20].

Abbate et al. [2] find that holes formed by A-site substitution go to states of primarily oxygen character, and the main component of the ground state in LSF50 is $3d^5\underline{L}$. In their study, and also in the XAS study on LSF by Wadati et al. [20], the $e_g$ peak grows with increasing Sr content and permits a quantitative consideration of hole doping by substitution at the the A-site.

Sarma et al. [1] have pointed to the situation where homovalent B-site substitution may cause a mismatch in the lattice, redistributing empty hole states near the Fermi level and thus improving the conductivity. We observe that substitution with the supravalent Ta and Ti cations appears to perform the opposite: blocking the conductivity.

In our study, where the B-site cation is substituted, we find that the change in the $e_g$ intensity alone does not permit systematic conclusions about the material, less the conductivity. We first tried to simply correlate the $e_g$ peak with the hole concentration $\tau$. But LSF50 and LSFTa10 differ by only about 5% in their $e_g$ intensity or peak area, whereas the conductivity differs by two orders of magnitude in the order LSF (50 S/cm) > Ti2 (2) > Ta1 (1) > Ta2 (0.1 S/cm). Hence, the peak intensity or peak area alone is not a good indicator for changes in the transport properties. Instead, as will be shown, it turns out that the ratio $e_g/t_{2g}$ correlates better with the hole concentration and the conductivity. De Groot et al. and Wu et al. [17,19] have elaborated on the effects that cause the relative intensities of these prepeaks in transition metal oxides, and conclude that the intensities are affected as follows in the order of significance: the number of 3d holes on Fe, hybridization effects, exchange interaction, and material non-stoichiometry.

We now show the correlation of conductivity and spectral weight versus the relative hole concentration $\tau$ in Figure 4. The abscissa on the left represents the conductivity on a logarithmic scale. The corresponding data points (open symbols) can be well fitted with an exponential of the form $\sigma = \sigma_0 \cdot \exp(A \cdot \tau)$. The conductivity thus decreases stronger than with any power of $\tau$, particularly not linearly, regardless whether we consider the formal (bottom axis) or the actual (top axis) hole concentration.

The abscissa on the right represents the ratio of the $e_g$ and $t_{2g}$ band area in the O(1s) spectra on the linear scale. The four data points show a linear correlation represented by a linear least square fit of the form $R(\tau)=R_0+ a \cdot \tau$.

The fact that we have a non-zero ratio $R_0$ for vanishing hole concentration requires an explanation. Certainly, LSFTa20 has some residual conductivity even in the absence of holes, i.e. the spectral intensity near the Fermi level and thus the corresponding conductivity is not only due to by hole states. Also, we cannot rule out that this intensity is caused by O(2p)-Ta(5d) mixed states. On the other hand, the unit cell is slightly expanding upon addition of Ta and Ti, which situation will generally not favor an increase in conductivity.

Altogether, the exponential conductivity decrease and the linear spectral weight decrease with decreasing hole concentration represent strong evidence that electron holes provide the majority of the transport, and that they can be monitored by the density of the unoccupied O(2p) states near the Fermi energy.

Tsipis et al. conclude that the total conductivity in Ti substituted LSF is lower than in LSF, because the more redox stable $Ti^{4+}$ ions lower the concentration of B-sites in electronic and hole transport. In addition do they believe that Ti addition does not alter any conductivity mechanism, but decreases the concentration and mobility of charge carriers [21].

Since the non-substituted LSF50 sample fits, too, in our proposed scheme, we believe that the $e_g/t_{2g}$ ratio as an indicator for transport properties is more general than for B-site substitution only. To further support this hypothesis, we have formed the $e_g/t_{2g}$ ratios from data presented by Wadati et al. [20], i.e LSF60, LSF20, as $LaFeO_3$ and plotted them in Figure 4 as well. These ratios can be expressed as a linear function of $\tau$ [here we have chosen the formal $\tau$, since the actual hole concentration was not available]. Note that we now have a set of A-site and B-site substituted LSF materials, which show a linear relation of $e_g/t_{2g}$ ratio and relative hole concentration: $R = 0.10 + 1.55 \cdot \tau$ (LSF-Ta), and $R=0.02+1.31 \cdot \tau$ (LSF, Wadati, [20]). In both cases do we have an increase of the unit cell volume [9]. By substitution of the B cation, rather than the A cation, we become more sensitive to the subtleties of the Fe-O-(Fe,Ta,Ti, ..) superexchange unit that ultimately provides – or blocks – the transport, particularly since the A cations usually do not participate in the transport.

However, we do not see why the $t_{2g}$ band, which is more remote from the Fermi energy than the $e_g$ band, should have any relevance to the transport properties. The $t_{2g}$ orbitals are localized and do not point towards the neighboring ligands, whereas the more delocalized $e_g$ bands do point to neighboring ligands and hence basically operate as switches for the charge transfer across the Fe-O-Fe superexchange unit. In this context it is interesting to note that the three samples with the highest conductivity also have the smaller branching ratios (Table).

$Fe^{4+}$ with $3d^4$ (according to Abbate [2] it is $3d5\underline{L}$) configuration assumes octahedral $d^2sp^3$ coordination, whereas $Fe^{3+}$ with $3d^5$ configuration assumes square/planar $dsp^2$ coordination. The depletion of $Fe^{4+}$ with increasing Ta or Ti substitution, and resulting enrichment with $Fe^{3+}$, causes rearrangement from octahedral to planar coordination, which consequently manifests in a depletion of the conducting $e_g$ bands, which require octahedral coordination.

Based on the observed exponential relation between hole concentration and conductivity, and the linear relation between hole concentration and ratio of the two bands $e_g/t_{2g}$, we propose the following mathematical relation between the conductivity and the band intensity ratio: $\sigma \sim \exp(A \cdot e_g/t_{2g})$, with A being a material specific constant.

Since the decrease of conductivity is exponential, but the hole variation and the spectral changes are linear, we believe that disorder and consequential percolation may be one primary cause for the conductivity decrease. Raychaudhuri [22] finds this to be the case for the systematic Ta doping on $Na_xWO_3$. Sarma et al. [23] come to similar conclusions with a systematic Ni/Fe substitution study on $LaNi_{1-x}Fe_xO_3$.

**Acknowledgement**

Financial support by the Swiss National Science Foundation, project # 200021-100674/1, and by the European Commission, contract # MIRG-CT-2006-042095 is gratefully acknowledged. The Science and Technology Facilities Council of the UK is acknowledged for granting and funding the beamtime at the Synchrotron Radiation Source under project # 47093.

| Stoichiometry<br>Label | δ | τ=$Fe^{4+}/Fe^{3+}$ | S/cm<br>@ 307 K | Q<br>$L_3/(L_2+L_3)$ | Exchange Splitting<br>Δ [eV] |
|---|---|---|---|---|---|
| $La_{0.5}Sr_{0.5}FeO_3$<br>LSF50 | 0.078 | 0.52 | 50 | 0.67 | 12.85 |
| $La_{0.5}Sr_{0.5}Fe_{0.8}Ti_{0.2}O_3$<br>LSFTi20 | 0.063 | 0.28 | 2.28 | 0.68 | 12.97 |
| $La_{0.5}Sr_{0.5}Fe_{0.9}Ta_{0.1}O_3$<br>LSFTa10 | 0.058 | 0.26 | 0.82 | 0.66 | 12.86 |
| $La_{0.5}Sr_{0.5}Fe_{0.8}Ta_{0.2}O_3$<br>LSFTa20 | 0.055 | 0.00 | 0.06 | 0.70 | 13.02 |

**Table:** Formal hole concentration, conductivity and branching ratio of $L_2$ vs. $L_3$.

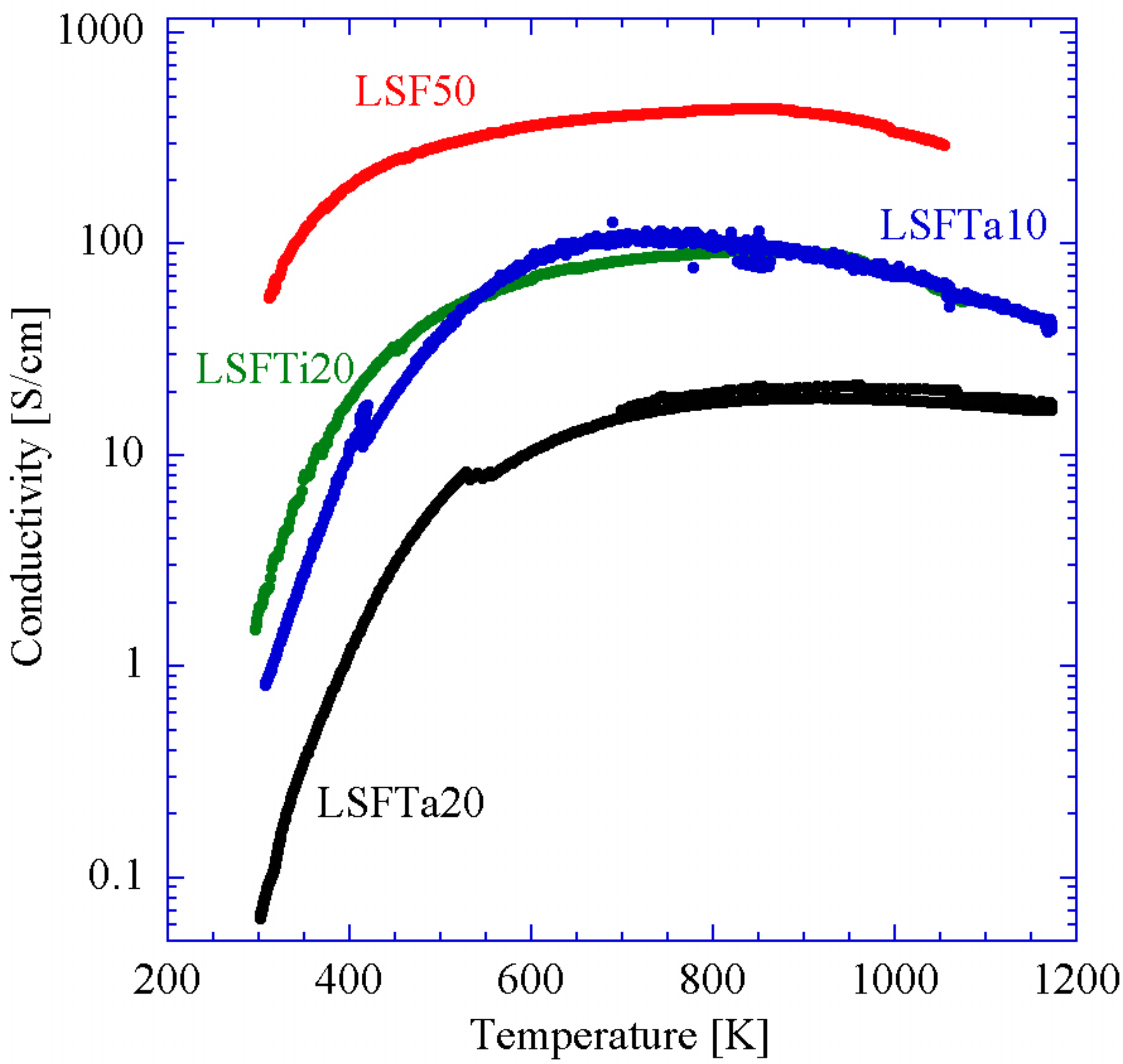

**Figure 1:** DC conductivity as a function of temperature as obtained by four-point technique.

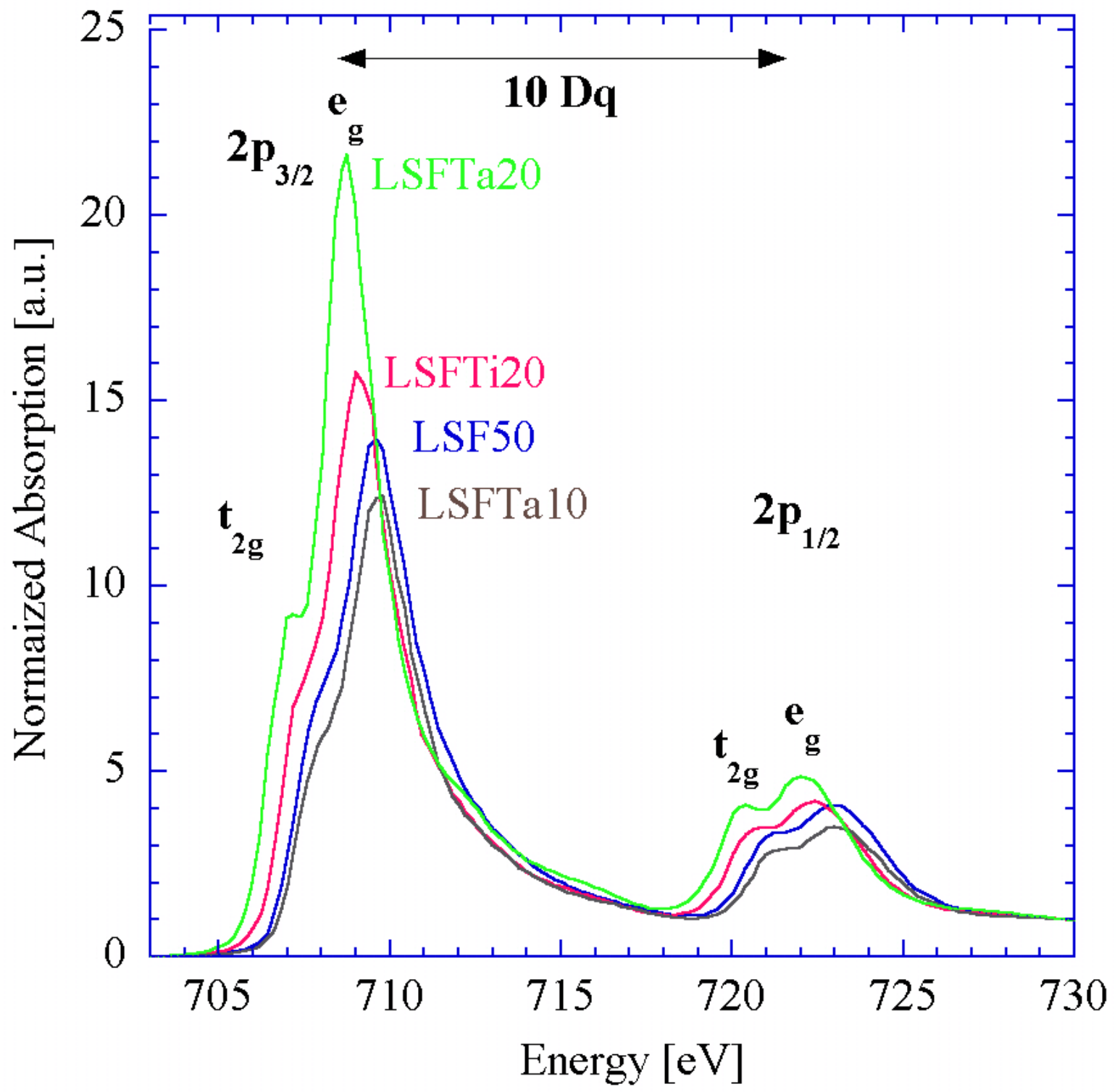

**Figure 2**: Fe L-edge absorption spectra.

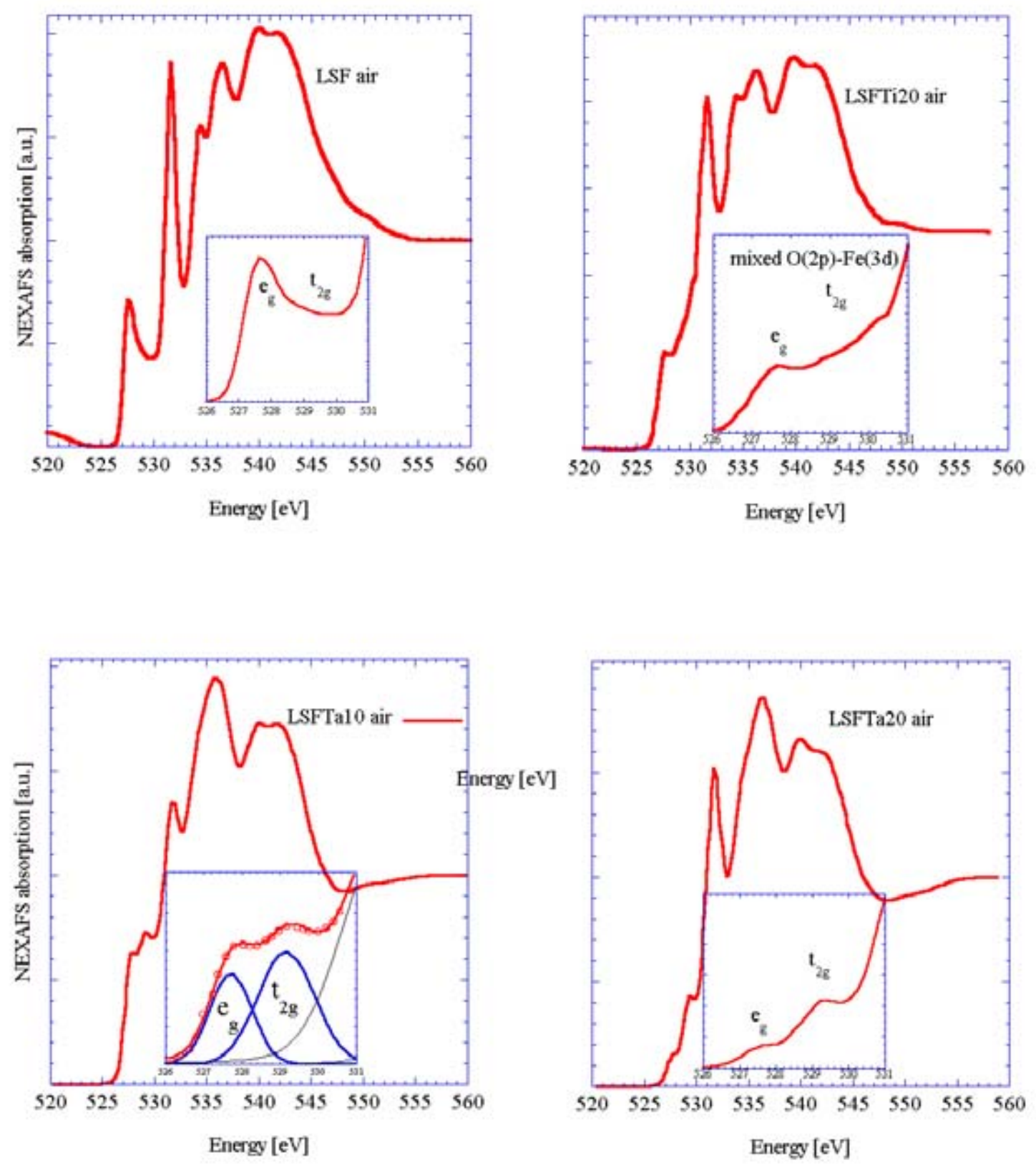

**Figure 3**: Oxygen (1s) absorption spectra. The insets show magnified areas near the Fermi level.

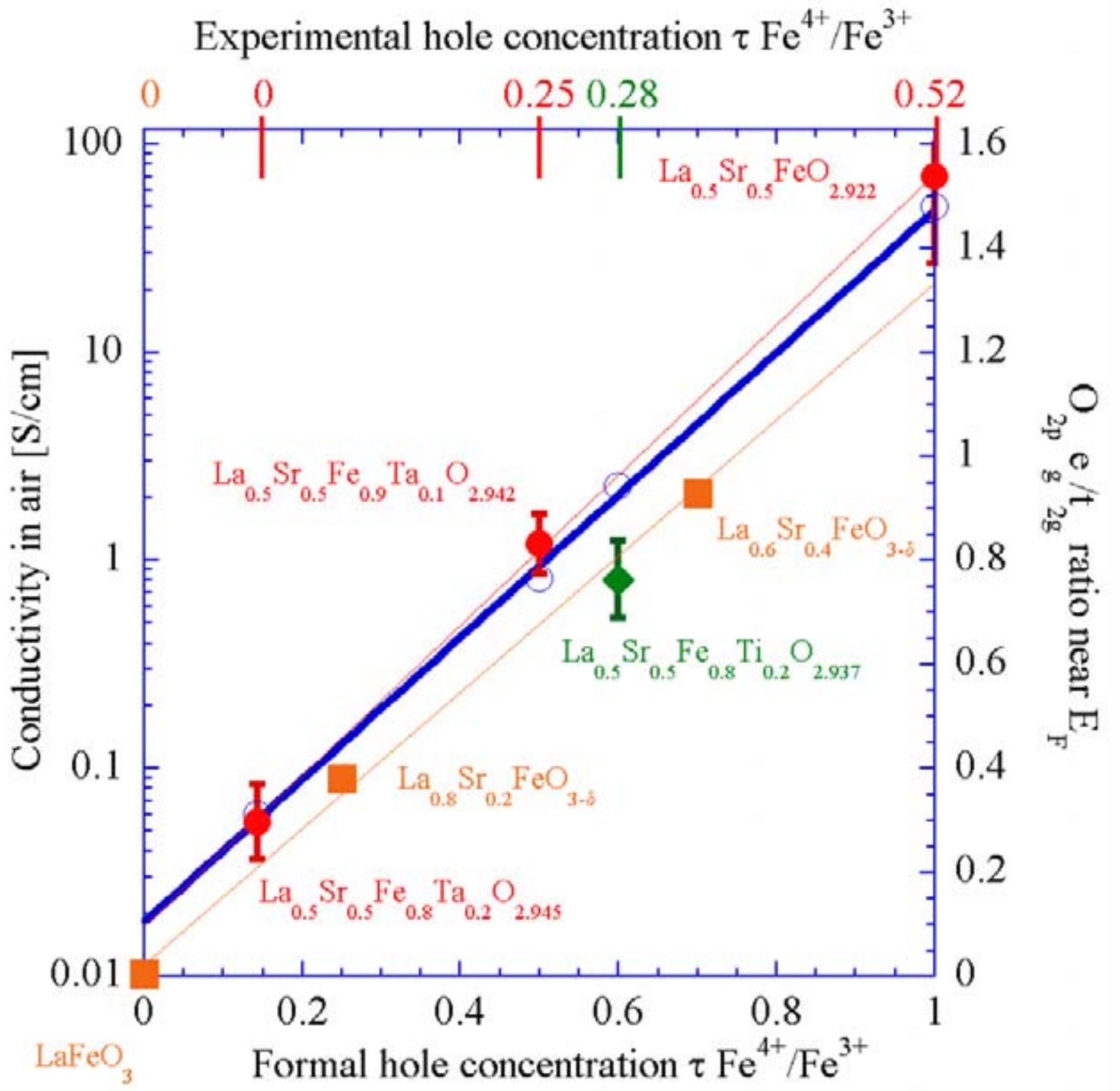

**Figure 4**: Correlation of Conductivity (open symbols with exponential fit, thick blue line), relative hole concentration and ratio of spectral intensity $e_g/t_{2g}$. The filled circles are from LSF50, LSFTa10 and LSFTa20, together with a linear least square fit line. The diamond is from LSFTi20. The squares are processed values taken from raw data by Wadati et al. [16], together with a least square fit line.